\documentclass[prd,twocolumn]{revtex4-2}
\usepackage{times}
\usepackage{amsmath}
\usepackage{amsfonts}
\usepackage{graphicx}
\usepackage[hidelinks]{hyperref}
\usepackage{dcolumn}
\usepackage{bm}

\def\beq{\begin{equation}}
\def\eeq{\end{equation}}
\def\bea{\begin{eqnarray}}
\def\eea{\end{eqnarray}}
\def\beqn{\begin{eqnarray}} 
\def\eeqn{\end{eqnarray}}
\def\beeq{\begin{eqnarray}}
\def\eeeq{\end{eqnarray}}

\def\ep{\epsilon}
\def\eps{\varepsilon}
\def\nn{\nonumber}
\def\Eq#1{Eq.~(\ref{#1})}

\def\qon#1{q_{#1,0}^{(+)}}

\def\res#1{{\rm Res}\left(#1\right)}

\def\qb{\mathbf{q}}
\def\pb{\mathbf{p}}
\def\kb{\mathbf{k}}

\def\lb{\boldsymbol{\ell}}

\def\uv{{\rm UV}}

\def\ii{\imath 0}

\def\F{{\rm F}}

\def\res#1{{\rm Res} \left(#1\right)}
\def\aa#1#2{{\cal A}_{(#2)}^{(#1)}}
\def\ad#1#2{{\cal A}_{{\rm D}(#2)}^{(#1)}}
\def\add#1#2#3{{\cal A}_{{\rm D}(#3)}^{(#1)|#2}}
\def\af#1#2{{\cal A}_{{\rm F}(#2)}^{(#1)}}

\begin{document}


\title{A Systematic Approach to Finite Multiloop Feynman Integrals}

\author{Prasanna K. Dhani~$^{(a)}$}\email{prasanna.dhani@physik.uzh.ch}
\author{Konstantinos Pyretzidis~$^{(c)}$}\email{k.pyretzidis@ific.uv.es}
\author{Selomit Ram\'{\i}rez-Uribe~$^{(b)}$}\email{selomitru@uas.edu.mx} 
\author{Jos\'e R\'{\i}os-S\'anchez~$^{(c)}$}\email{jose.rios@ific.uv.es}
\author{German F.R. Sborlini~$^{(d)}$}\email{german.sborlini@usal.es} 
\author{Surabhi Tiwari~$^{(c)}$}\email{surabhi.tiwari@ific.uv.es} 
\author{Germ\'an Rodrigo~$^{(c)}$}\email{german.rodrigo@csic.es}
\affiliation{
${}^{a}$ Physik Institut, Universit\"at Z\"urich, Winterthurerstrasse 190, CH-8057 Z\"urich, Switzerland. \\
${}^{b}$ Facultad de Ciencias F\'{\i}sico-Matem\'aticas, Universidad Aut\'onoma de Sinaloa, Ciudad Universitaria, CP 80000 Culiac\'an, Mexico. \\
${}^{c}$ Instituto de F\'{\i}sica Corpuscular, Universitat de Val\`{e}ncia -- Consejo Superior de Investigaciones Cient\'{\i}ficas, Parc Cient\'{\i}fic, E-46980 Paterna, Valencia, Spain. \\
${}^{d}$ Departamento de F\'isica Fundamental e IUFFyM, Universidad de Salamanca, 37008 Salamanca, Spain.
}

\date{March 19, 2026}

\begin{abstract}
Finite Feynman integrals have been advocated as the optimal components for constructing a basis of master integrals in multiloop calculations, due to their improved analytic and numerical properties. In this paper, we show how the Loop-Tree Duality (LTD) is particularly well suited for systematically identifying finite integrals, as it makes the origin of infrared and threshold singularities fully transparent at the integrand level. This clear separation of singular and non-singular contributions enables a more efficient strategy for isolating and promoting finite integrals, thereby streamlining both reduction and numerical evaluation. We present a new strategy based on numerator and raised propagator Ans\"atze that provides results similar to other methods, although in a clearer and compact way. While this construction and other approaches establish a robust foundation, they often produce integrands that exhibit a rapid growth in the ultraviolet (UV) regime. To mitigate this bad UV behaviour, we introduce a generalized set of integrands fully defined within LTD. This new set is inherently infrared-finite and frequently free of threshold singularities, offering a more versatile framework for high-order calculations.
\end{abstract}

\maketitle

\section{Introduction}

The identification of a basis of finite Feynman integrals has recently emerged as a powerful strategy for constructing an optimal set of master integrals and enhancing the numerical implementation of multiloop calculations in Quantum Field Theory (QFT). The standard approach to handling infrared~(IR) divergences in QFT usually involves sector decomposition techniques~\cite{Hepp:1966eg, Roth:1996pd,Binoth:2000ps,Binoth:2003ak,Bogner:2007cr}, with several public codes available, such as  {\tt pySecDec}~\cite{Borowka:2017idc} and {\tt Fiesta5}~\cite{Smirnov:2021rhf}, that automate this procedure. However, these methods can become computationally challenging for complex topologies. On the contrary, IR-finite integrals exhibit enhanced numerical stability, as IR poles are absorbed into coefficients, greatly simplifying the physical four-dimensional limit $d \to 4$~\cite{Borowka:2016ypz, vonManteuffel:2014qoa}.

An explicit construction of finite integrals can be achieved through a dimensional shift~\cite{Bern:1992nf}, or by establishing a finite basis in the Euclidean region~\cite{Panzer:2014gra, vonManteuffel:2014qoa}. Tools such as {\tt Reduze2}~\cite{vonManteuffel:2012np, Studerus:2009ye} enable the identification of finite integrals by introducing both raised propagators (so-called “dots”) and dimensional shifts. Nonetheless, integrals with many dots are often difficult to reduce and may not be necessary for amplitude reduction. Moreover, higher powers in the denominators degrade the performance of contour deformation~\cite{Jones:2025jzc}. 

A more efficient approach involves using suitable numerators to cancel IR singularities directly at the integrand level. Recent developments focus on constructing linear combinations of individually divergent integrals whose combined numerators produce finite integrands in four spacetime dimensions. This method leverages structures already present in scattering amplitudes, avoiding the need for additional reductions and providing a more natural $d=4$ representation. Nevertheless, identifying the numerators of such a set of integrals remains a non-trivial challenge.
For example, the {\tt FineComb} algorithm~\cite{Agarwal:2020dye} generates all possible finite linear combinations from a given set of seed integrals. The algorithm merges integrals from the same parent topology combining numerators and subsectors into a unified Feynman parametric representation and inspects their scaling behavior near each boundary. Divergences are eliminated by imposing vanishing constraints on offending terms. These constraints, solvable for instance with {\tt FiniteFlow}~\cite{Peraro:2019svx}, yield finite integrands. Although the method is computationally demanding due to the complexity of the merged denominator and numerator structures, {\tt FineComb} implements optimizations that significantly improve efficiency.

Other approaches, such as those based on Landau analysis~\cite{Gambuti:2023eqh} and Newton polytopes~\cite{delaCruz:2024xsm}, further advance the identification of finite integrands. The Landau equations enable the study of singularity structures by analyzing the configurations of loop momenta that give rise to potential divergences. Meanwhile, the Newton polytope provides a complementary perspective by encoding the analytic and combinatorial properties of integrands into convex geometric objects, elucidating the connections between algebraic geometry and the analytic behavior of scattering amplitudes. Together, these methods deepen our understanding of the geometric foundations of IR-finite constructions, revealing patterns and symmetries that may guide novel finiteness criteria. Further developments involve an initial set of finite integrals to iteratively  construct an Integration-By-Parts~(IBP) basis in which singularities are absent in the reduction coefficients~\cite{DeAngelis:2025agn}.

In this Letter, we leverage the Loop-Tree Duality (LTD)~\cite{Ramirez-Uribe:2024rjg,Aguilera-Verdugo:2020set,Capatti:2019ypt,Runkel:2019yrs,Bierenbaum:2012th,Bierenbaum:2010cy,Catani:2008xa,Soper:1998ye} to explore a more direct and streamlined framework for classifying IR-finite multiloop Feynman integrands. Three different strategies are defined, with the third one, intrinsically based on LTD, showing a milder scaling of ultraviolet (UV) singularities compared to integrands built in the Feynman representation. In most cases, these integrands are naturally free of threshold singularities when considering physical kinematics. This inherent property simplifies numerical implementations by removing the need for additional contour deformation~\cite{Soper:1999xk,Binoth:2000ps,Binoth:2005ff,Nagy:2006xy}, threshold subtractions~\cite{Kermanschah:2024utt,Kermanschah:2021wbk}, extrapolation~\cite{deDoncker:2017gnb,deDoncker:2004fb}, $\imath 0$-deformation~\cite{Pittau:2021jbs} or shifts to the Euclidean region.

\section{Multiloop Feynman integrals in the Feynman and LTD representations}

A multiloop Feynman integral in the Feynman representation is generally described as 
\beq
\aa{\Lambda}{a_1,\ldots,a_n} = \int_{\ell_1 \cdots \ell_\Lambda} \af{\Lambda}{a_1,\ldots,a_n}~,
\label{eq:feynman}
\eeq
where $\{\ell_i\}_\Lambda$ are $\Lambda$ independent loop four-momenta, and the integration measure in $d$-spacetime dimensions is 
$\int_{\ell_s} = -\imath \mu^{4-d} \int d^d\ell_s/(2\pi)^d$.
The integrand in~\Eq{eq:feynman} is further specified as 
\beq
\af{\Lambda}{a_1,\ldots,a_n} = {\cal N}^{(\Lambda)}_{(a_1,\ldots,a_n)}(\{\ell_i\}) \, I_{\F (a_1,\ldots,a_n)}^{(\Lambda)}~,
\eeq
where ${\cal N}^{(\Lambda)}_{(a_1,\ldots,a_n)} (\{\ell_i\})$ is a numerator function, the dependence on the external momenta is understood, and the scalar function 
\beq
I_{\F (a_1,\ldots,a_n)}^{(\Lambda)} = \prod_{i=1}^n \left( G_{\F}(q_i) \right)^{a_i}~,
\eeq
is written in terms of Feynman propagators, where $G_{\F}(q_i) = (q_i^2-m_i^2+\ii)^{-1}$. The momenta of the Feynman propagators, $q_i$, are expressed as linear combinations of the loop and external momenta. We assume that the numerator does not have any open tensor index, and is written in terms of scalar products of loop and external momenta. Alternatively, the numerator can be rewritten in terms of the inverses of the Feynman propagators, where the power coefficients $\{a_i\}_n$ take on negative values, assuming there is a complete basis of denominators. 

LTD reformulates the multiloop Feynman integral in \Eq{eq:feynman} by using Cauchy's residue theorem to integrate over one component of each loop momenta, effectively changing the integration domain, and generating a dual representation which is manifestly causal. If the energy components are integrated out, the loop integral has support in the Euclidean space of the loop three-momenta and the multiloop integral takes the form
\beq
\aa{\Lambda}{a_1,\ldots,a_n} = \int_{\lb_1 \cdots \lb_\Lambda} 
\ad{\Lambda}{a_1,\ldots,a_n}~,
\eeq
where 
$\int_{\lb_s} = \mu^{4-d} \int d^{d-1}\lb_s/(2\pi)^{d-1}$ is the integration measure,
and $\ad{\Lambda}{a_1,\ldots, a_n}$ the LTD representation of $\af{\Lambda}{a_1,\ldots, a_n}$. This LTD representation is a function of the on-shell energies $\qon{i} = \sqrt{\qb_{i}^2+m_{i}^2-\ii}$, where $\qb_i$ are the spacial components of the propagators momenta and the $\ii$-prescription stems from the original complex prescription of the Feynman propagators. Specifically, in the LTD framework, the Feynman propagators are replaced by causal propagators, which are expressed as linear combinations of the internal on-shell energies and the energies of the external particles.

\section{Collinear and soft singularities}

In LTD, a double collinear singularity is encoded by a causal propagator of the form $1/\lambda_{i_3 i_2 \bar i_1}$~\cite{Imaz:2025buf,LTD:2024yrb,Kromin:2022txz,Sborlini:2021owe,TorresBobadilla:2021ivx,Aguilera-Verdugo:2019kbz,Buchta:2014dfa}, with
\beq
\lambda_{i_3 i_2; \bar i_1} = \qon{i_3}+\qon{i_2}-p_{i_1,0}~,
\eeq
where $p_{i_1}$ is a massless outgoing external particle, $p_{i_1}^2=0$, with $p_{i_1,0}=|\pb_{i_1}|>0$. For incoming external particles, we use the convention  $p_{i_1,0}<0$ and a collinear singularity is encoded by $\lambda_{i_3 i_2; i_1} = \qon{i_3}+\qon{i_2}+p_{i_1,0}\to 0$. The internal particles must also be massless and adjacent, i.e. $q_{i_3} + q_{i_2} = p_{i_1}$. The internal particles are generally off-shell, and the singularity is activated when $q_{i_s,0} = \qon{i_s} = \sqrt{\qb_{i_s}^2-\ii}$, i.e. for on-shell configurations. 

As a novelty, we introduce a  collinear parametrization in the three-momenta
\beq
\qb_{i_s} = z_{i_s} \pb_{i_1} + \kb_{i_s,\perp}~,  \quad s=2,3~,
\eeq
with $\kb_{i_s,\perp}\cdot \pb_{i_1} = 0$. Note that unlike the customary Sudakov parametrization, no auxiliary lightlike vector is required here to specify the collinear direction and enforce the on-shell condition of the collinear particles. In the limit of small transverse momentum~\footnote{The actual collinear scaling is $\lambda_{i_3 i_2; \bar i_1} \sim k_{i_2,\perp}^2/|\pb_{i_1}|$}
\beq
\lambda_{i_3 i_2; \bar i_1} = {\cal O} (k_{i_2,\perp}^2)~,
\eeq
assuming a symmetric configuration with $\kb_{i_3,\perp} = - \kb_{i_2,\perp}$, and $k_{i_2,\perp}^2=\kb_{i_2,\perp}^2$.

The generalization to configurations involving multiple collinear particles is 
\beq
\lambda_{i_r \cdots i_2; \bar i_1} = \sum_{s=2}^r \qon{i_s} - p_{i_1,0} = {\cal O} (k_{i_s,\perp}^2)~,
\eeq
where the internal particles are adjacent, in the sense that they satisfy $\sum_{s=2}^r q_{i_s} = p_{i_1}$.

A soft singularity arises from two adjacent causal propagators, such as $1/\lambda_{i_3 i_2; \bar i_1}$ and $1/\lambda_{i_3 i_4; \bar i_5}$, where $q_{i_3} + q_{i_2} = p_{i_1}$ and $q_{i_3} + q_{i_4} = p_{i_5}$. So, by introducing the scaling $\qb_{i_3} \to \rho \,\qb_{i_3}$ and examining the limit $\rho \to 0$, each causal propagator scales as ${\cal O}(\rho^{-1})$, leading to a $1/\rho^2$ quadratic singular behavior near the soft configuration. Simultaneously, the integration measure in $d-1=3-2\eps$ spacial dimensions and a factor $1/\qon{i_3}$ from the LTD representation contribute to a factor of $\rho^2$, leading to the expected logarithmic singularity at $\rho \to 0$. 

\section{Methodology}

Our methodology consists in exploiting the LTD representation of a target integrand by evaluating residues corresponding to collinear configurations. The numerator of this target integrand or Ansatz is constructed as a function of a number of arbitrary coefficients. LTD transforms this numerator, initially defined in the Feynman representation, into a polynomial in the on-shell energies of the internal particles that depends on the arbitrary coefficients given by the original numerator. We then impose that the collinear residues vanish,  
\beq
\res{\ad{\Lambda}{a_1,\ldots, a_n}, \lambda_{i_r\cdots i_2; \bar i_1}} = 0~.
\eeq
Assuming that each monomial in the on-shell energies is independent, we set the factors of each term to zero. This procedure generates a system of linear equations in the coefficients of the numerator, which we solve.

If a Feynman diagram contains propagators raised to a given power, then the corresponding LTD representation will contain causal propagators raised to a power, together with subleading terms in the causal propagators. In general, in the collinear limit
\beq
\left(\lambda_{i_r \cdots i_2; \bar i_1}\right)^a = {\cal O} (k_{i_s,\perp}^{2a})~.
\eeq
For instance, in $d+2$ spacetime dimensions, with $d=4-2\ep$, a term of the type $1/(\lambda_{i_r\cdots i_2; \bar i_1})^2$ is IR singular, but the subleading term $1/\lambda_{i_r\cdots i_2; \bar i_1}$ is IR finite. So, another set of IR-finite integrands with raised propagators is determined by imposing 
\beq
\res{\left(\lambda_{i_r\cdots i_2; \bar i_1}\right)^{a-1} \ad{\Lambda}{a_1,\cdots, a_n},  \lambda_{i_r\cdots i_2; \bar i_1}} = 0~,
\eeq
and shifting the spacetime dimensions of the loop integral to $d+a$. 

Regarding soft singularities, enforcing the cancellation of the residues of two adjacent causal propagators effectively removes the soft singularity. In fact, in the absence of one of the two collinear causal propagators the potential soft singularity is suppressed by the integration measure. In $d+a$ spacetime dimensions, the integration measure scales as $\rho^{d+a-1}$, leading to a faster suppression in the soft limit. While raised propagators may compensate for the soft behaviour of the integration measure, soft singularities are generally accounted for as a byproduct of suppressing collinear configurations.

The kinds of terms we expect to pass these tests are as follows. A double collinear singularity, $\lambda_{i_3 i_2; \bar i_1}\to 0$, occurs when one internal particle is on-shell and a second, neighbouring one, is approaching its on-shell state, i.e., a term in the numerator such that 
\beq
\left. q_{i_3}^2\right|_{q_{i_2}^2, p_{i_1}^2=0} = \left. (p_{i_1}-q_{i_2})^2 \right|_{q_{i_2}^2, p_{i_1}^2=0} =  -2 q_{i_2}\cdot p_{i_1}~,
\label{eq:doublecollinear}
\eeq
would cancel the collinear singularity that arises as $q_{i_3}^2\to 0$, when $q_{i_2}$ is already on shell. Note that the symmetric configuration where $q_{i_3}^2=0$ and $q_{i_2}$ approaches its on-shell state would be equivalent because of the adjacency condition, i.e. $q_{i_2}\cdot p_{i_1} = -q_{i_3}\cdot p_{i_1}$. Note that this term can be rewritten as 
\beq
-2q_{i_2}\cdot p_{i_1}= q_{i_3}^2-q_{i_2}^2~.
\eeq
Another possibility involves removing the internal propagators that connect two external massless particles, thereby generating an effective interaction vertex with both external particles attached. Terms of this kind take the form
\beq
{\cal N}^{(\Lambda)}_{(a_1, \ldots, a_n)} (\{\ell_s\}) \propto \prod_r q_{i_r}^2~,
\label{eq:effective}
\eeq
where the product runs over the relevant momenta. A numerator factor formed by the product of contributions from \Eq{eq:doublecollinear} and \Eq{eq:effective} would similarly yield IR-finite integrands. However, the construction of numerators with exclusively terms of the type in \Eq{eq:doublecollinear} and \Eq{eq:effective} is restricted if we impose UV finiteness.

\section{Finite integrals at one-loop}

\begin{figure}[t]
\begin{center}
\includegraphics[scale=.58]{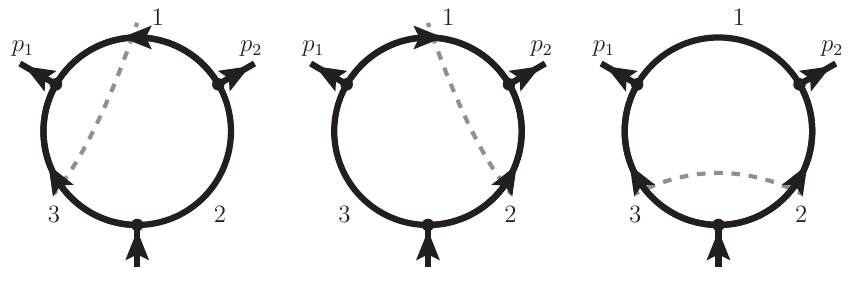}
\caption{Collinear and threshold configurations of the one-loop three-point function. From left to right, represented by $1/\lambda_{31;\bar 1}$, $1/\lambda_{12;\bar 2}$ and $1/\lambda_{23;\overline{12}}$. The mirror configurations, with the internal particles propagating in the opposite directions, are represented by $1/\lambda_{31;1}$, $1/\lambda_{12;2}$ and $1/\lambda_{23;12}$, and are collinear finite and threshold free. 
\label{fig:oneloop}}
\end{center}
\end{figure}

We illustrate the method with a few examples that feature the main properties. We consider a three-point integrand at one-loop (Fig.~\ref{fig:oneloop}), $\aa{1}{a_1,a_2,a_3}$, with the definition of the internal momenta $q_i = \ell_1 + k_i$, $i = 1, 2, 3$, where $k_1 = p_1$, $k_2 = p_1+p_2 = p_{12}$ and $k_3 = 0$. The third external momentum is fixed by momentum conservation, $p_3^2 = p_{12}^2 = s.$ All the internal particles are massless, and $p_1^2=0$. If $p_2^2 \ne 0$, the only IR singularity occurs at $\lambda_{31;\bar 1} \to 0$.

Assuming a numerator of the form
\beq
{\cal N}^{(1)}_{(1,1,1)}(\ell_1) = \alpha_0 + \alpha_1 \, \ell_1\cdot k_1 + \alpha_2 \, \ell_1\cdot k_2~,
\eeq
we obtain the corresponding LTD representation and impose
\beq
\res{\ad{1}{1,1,1},\lambda_{31;\bar1}} = 0~.
\eeq
This condition gives as a result $\alpha_0 = \alpha_2 = 0$, or ${\cal N}_{(1,1,1)}^{(1)}(\ell_1) \sim \ell_1\cdot p_1  = q_3\cdot p_1$.
If instead $p_2^2=0$ and $p_1^2 = m_1^2 > 0$, the solution is $\alpha_1=-\alpha_2$ and $\alpha_0 = \alpha_2 (s-m_1^2)/2$, i.e.
\beq
{\cal N}^{(1)}_{(1,1,1)}(\ell_1) \sim (\ell_1+p_1)\cdot p_2 = q_1\cdot p_2~, 
\label{eq:theother}
\eeq
which is obtained by setting the residue at $\lambda_{12;\bar 2}$ to zero.

If $p_1^2=p_2^2=0$, we should set to zero the two residues simultaneously. A linear numerator in the loop momentum gives a trivial solution, so we extend the numerator to a polynomial of rank $2$, which is UV singular,
\bea
{\cal N}^{(1)}_{(1,1,1)}(\ell_1) &=& \alpha_0 + \alpha_1 \, \ell_1\cdot k_1 + \alpha_2 \, \ell_1\cdot k_2 \nn \\ &+& \alpha_3 \, (\ell_1\cdot k_1)(\ell_1\cdot k_2) + \alpha_4 \, (\ell_1\cdot k_1)^2 \nn \\ &+& \alpha_5 \, (\ell_1\cdot k_2)^2 + \alpha_6 \, \ell_1^2~.
\eea
If we now impose that the residue at $\lambda_{31;\bar 1}$ and $\lambda_{12;\bar 2}$ vanish simultaneously, we find the following solution in terms of two independent coefficients (the coefficients are conveniently rescaled and relabeled because they are arbitrary)
\bea
{\cal N}^{(1,\uv) | d}_{(1,1,1)} &=& \alpha_0 \left( q_1^2-q_3^2\right)\left(q_1^2-q_2^2\right) + \, \alpha_1 \, q_1^2 \nn \\
&=& \alpha_0 (q_1\cdot p_1)(q_2\cdot p_2)+ \alpha_1 \, q_1^2~. 
\label{eq:num111}
\eea
The index $\uv$ indicates that the corresponding integrand is UV singular, although IR finite. We also specify that the integral is defined in $d=4-2\eps$ spacetime dimensions. These findings align with the general discussion outlined above.

Let's now analyze the case of a raised propagator, e.g. $\aa{1}{2,1,1}$. In $d+2$ spacetime dimensions, this integrand is UV finite for a linear numerator. However, the first nontrivial IR finite integrand appears for a rank 2 numerator 
\bea
{\cal N}^{(1,\uv) | d+2}_{(2,1,1)} &=& {\cal N}^{(1,\uv) | d}_{(1,1,1)}~.
\label{eq:num111d6}
\eea
In $d=4-2\epsilon$, a numerator of rank 3 is UV finite, and an IR safe integrand is obtained with 
\bea
{\cal N}^{(1) | d}_{(2,1,1)} &=& \alpha_0 \, \left(q_1^4-q_2^2 q_3^2\right)~.
\label{eq:num211}
\eea
It is interesting to note that from the general power-counting arguments discussed above, one would instead anticipate $(q_1\cdot p_1)^2(q_2\cdot p_2)^2$, which is rank $4$, and therefore UV.

If the raised propagator is the third one, we find instead
\bea
{\cal N}^{(1) | d}_{(1,1,2)} &=& \alpha_0 \, (q_1\cdot p_1)^2(q_2\cdot p_2) + r_1 \, q_1^2~, \nn \\
{\cal N}^{(1) | d+2}_{(1,1,2)} &=&  \alpha_0 \, (q_1\cdot p_1)~.
\eea
While if two propagators are raised
\bea
{\cal N}^{(1) | d}_{(1,2,2)} &=&  r_1 \, (q_1\cdot p_1)^2(q_2\cdot p_2)^2 + r_3 \, q_1^2~, \nn \\
{\cal N}^{(1) | d+2}_{(1,2,2)} &=&  r_1 \, (q_1\cdot p_1)(q_2\cdot p_2) + r_1' \, q_1^2~, \nn \\
{\cal N}^{(1) | d+4}_{(1,2,2)} &=& r_1~, 
\eea
where $r_i$ is a polynomial of rank $i$ in the loop momentum, e.g., $r_1$ ($r_1'$) is a linear numerator, and $r_3$ is a cubic numerator. They contain $3$ and $13$ independent terms, respectively. Once again, the pattern is different if the first propagator is raised, 
\bea
{\cal N}^{(1) | d}_{(2,1,2)} &=& \left( \alpha_0 \, (q_1\cdot p_1)^2(q_2\cdot p_2)^2 + r_2 \, q_1^2 \right) (q_1\cdot p_1)~, \nn \\
{\cal N}^{(1) | d+2}_{(2,1,2)} &=&  {\cal N}^{(1,\uv) | d}_{(1,1,1)} \, (q_1\cdot p_1)~, \nn \\
{\cal N}^{(1) | d+4}_{(2,1,2)} &=& \alpha_0 \, (q_1\cdot p_1)~.
\eea
Finally, when the three propagators are raised
\beq
{\cal N}^{(1) | d+4}_{(2,2,2)} =  {\cal N}^{(1,\uv) | d}_{(1,1,1)}~, \qquad
{\cal N}^{(1) | d+6}_{(2,2,2)} =  r_1~. 
\eeq

\section{Two-loop triangle}

\begin{figure}[t]
\begin{center}
\includegraphics[scale=.58]{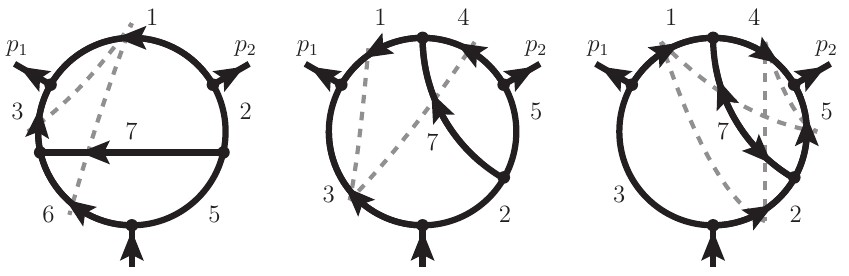}
\caption{Collinear configurations of the two-loop Feynman diagrams with three external massless particles.  The direction of propagation of particle $7$ in the third diagram depends on the causal configuration considered.
\label{fig:twoloop}}
\end{center}
\end{figure}

\begin{figure}[t]
\begin{center}
\includegraphics[scale=.68]{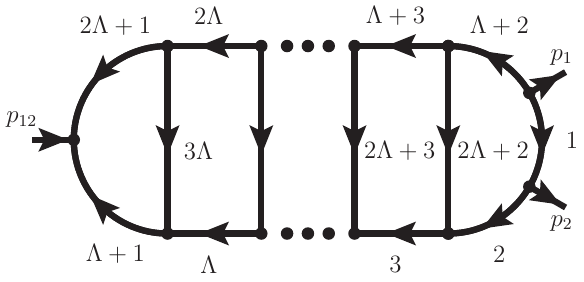}
\caption{IR-finite and threshold-free multiloop ladder configuration with three external massless particles.
\label{fig:ladder_triangle}}
\end{center}
\end{figure}

The two-loop Feynman integrands that we analyze are the three-point functions in Fig.~\ref{fig:twoloop}, with internal propagators defined by 
$q_{i} = \ell_1 + k_i$, 
$q_{3+i} = \ell_2 - k_i$, 
$q_{7} = \ell_1 + \ell_2$, 
where $i = 1, 2, 3$, with $k_1=p_1$, $k_2=p_{12}$ and $k_3=0$. Our first target integrand is $\aa{2}{1,1,1,0,1,1,1}$. Note that there are two residues corresponding to double-collinear configurations and another two corresponding to tripple-collinear configurations. A numerator proportional to $(q_1\cdot p_1)(q_2\cdot p_2)(q_1\cdot q_5)(q_3\cdot q_4)$ would cancel the four collinear residues, but it is of rank $4$ in $\ell_1$ and rank $2$ in $\ell_2$. Instead, we test a numerator of rank $(3,1)$, which is UV finite, and find
\beq
{\cal N}^{(2) | d}_{(1,1,1,0,1,1,1)} = r_{1,1} \, q_1^2~,
\eeq
where $r_{1,1}$ is a polynomial of rank $1$ in both $\ell_1$ and $\ell_2$ consisting of $10$ independent terms. 

As a second example, we consider $\aa{2}{1,1,1,1,1,0,1}$. However, a numerator of rank (3,1) gives a trivial result because there are six collinear residues that introduce too many constraints. A numerator of rank (3,2) is IR finite, although UV singular in $\ell_2$,
\beq
{\cal N}^{(2,\uv) | d}_{(1,1,1,1,1,0,1)} = r_{1,0} \, q_1^2 q_4^2~.
\label{eq:num11111010}
\eeq

\section{A new set of IR-finite and threshold-free integrals}

In the previous section, we considered numerators constructed from scalar products involving loop and external four-momenta. We then imposed the condition that the residues on causal propagators, $1/\lambda_{i_r\cdots i_2; \bar i_1}$, vanish to adjust the coefficients of these numerators and suppress the IR singularities. Another possibility is obtained by constructing numerators directly in terms of on-shell energies. Following the method described previously, we depart from a general Ansatz with arbitrary coefficients and impose that certain residues vanish. It is not surprising that this construction aligns the coefficients of the on-shell energies in such a way that the resulting terms correspond to the product of all causal denominators associated with IR singularities. For example, 
\bea
&& {\cal N}^{(1,\uv) | d}_{(1,1,1)} = {\cal N}^{(1,\uv) | d+2}_{(2,1,1)} = \alpha_0 \, \lambda_{31;\bar 1} \lambda_{12;\bar 2}~, \nn \\
&& {\cal N}^{(1,\uv) | d}_{(2,1,1)} = \alpha_0 \, \lambda_{31;\bar 1}^2 \lambda_{12;\bar 2}^2~.
\eea
These terms play the same role as the numerators in \Eq{eq:num111}, \Eq{eq:num111d6}, and \Eq{eq:num211}, with the important distinction that their dependence on the energy component of the loop momenta is absent. 

All these integrands, while IR finite, exhibit in most cases UV singularities. However, it is important to recognize that the two factors in $(q_1\cdot p_1)(q_2\cdot p_2)$ or $\lambda_{31;\bar 1} \lambda_{12;\bar 2}$ vanish simultaneously only in the soft limit, while each collinear configuration remains independent. As a result, we observe that a multiplicative construction leads to numerators that grow rapidly in the UV regime, enhancing the UV behavior of the integrands. 

LTD offers a natural solution to define IR finite integrands while limiting the growth of UV singularities. The LTD representation is given by a sum of terms composed of products of causal propagators. Specifically, the maximum number of causal propagators in each term is given by the difference between the number of internal particles and the number of loops. For scalar integrals, all the terms contain exactly the maximum number of causal propagators, while for tensor integrals the presence of a numerator reduces this number for some contributions. However, it is important to note that IR singularities are encoded only in few subsets of terms. So, terms encoding specific configurations are IR finite without requiring additional numerators that could adversely affect the UV behaviour. 

Each causal propagator fixes the direction of propagation of the internal particles that are involved in that causal propagator. Thus, causal propagators in this context play a key role in fixing the direction of propagation of internal particles. Each causal propagator enforces an aligned causal direction in the flow of virtual particles.  This avoids contributions from nonphysical singularities that appear in conventional loop integrands. Depending on whether the momenta of the internal particles are also aligned with the external momenta, collinear and threshold singularities emerge. Causality constraints imply that the corresponding physical configurations are acyclic when regarded as directed graphs~\cite{Ochoa-Oregon:2025opz,Ramirez-Uribe:2024wua,Clemente:2022nll,Ramirez-Uribe:2021ubp,Ramirez-Uribe:2020hes,Aguilera-Verdugo:2020kzc}.

For the one-loop three-point function, we find the following family of IR- and UV-finite integrands
\beq
\add{1}{d}{1,1,1} = 
\frac{1}{x_{123}} \left(
\frac{r_1}{\lambda_{31;1}} +
\frac{r_1'}{\lambda_{12;2}}  \right) \frac{1}{\lambda_{23;12}} ~,
\label{eq:oneloop}
\eeq
where $r_1$ and $r_1'$ are linear numerators in the on-shell energies, and $x_{123} = \prod_{s=1}^3 2\qon{s}$.
These integrands are also threshold free because the causal denominator 
\beq
\lambda_{23;12} = \qon{2} + \qon{3} + p_{12,0}~,
\eeq
cannot vanish, assuming $p_{12,0}>0$, contrary to $\lambda_{23;\overline{12}} = \qon{2} + \qon{3} - p_{12,0}$, which is absent because it is associated to the collinear causal propagators. The corresponding collinear and threshold configurations are shown in Fig.~\ref{fig:oneloop}, which are absent in \Eq{eq:oneloop}.

At two loops there are several topologies with three external particles allowed, whose collinear configurations are shown in Fig.~\ref{fig:twoloop}. As illustrative examples of finite integrands, we consider the following two expressions
\bea
&& \add{2}{d}{1,1,1,0,1,1,1} = \frac{1}{x_{123567}} \frac{1}{\lambda_{13;1} \lambda_{167;1} \lambda_{267;12} \lambda_{56;12}
}~, \nn \\
&& \add{2}{d}{1,1,1,1,1,0,1} = \frac{1}{x_{123457}} \frac{1}{\lambda_{13;1} \lambda_{347;1} \lambda_{357;12} \lambda_{32;12}
}~, \nn \\
\label{eq:twoloops}
\eea
which correspond to noncollinear configurations of the two-loop diagrams in Fig.~\ref{fig:twoloop}. Here, $x_{i_1\cdots i_n} = \prod_{s=1}^n 2\qon{i_s}$. We also consider the ladder diagram depicted in Fig.~\ref{fig:ladder_triangle} to all loop orders, $\add{\Lambda}{d}{1,\cdots, 3\Lambda}$. Details on the loop labeling and the specific finite configuration considered are given in Appendix~\ref{app:ladder}. 

For benchmarking, we have implemented the finite integrands in~\Eq{eq:oneloop}, \Eq{eq:twoloops}, and  Fig.~\ref{fig:ladder_triangle} at different loop orders, by using VEGAS~\cite{Lepage:1977sw,Hahn:2004fe}. Numerical results are summarized in Table~\ref{tab:numerics}.

\begin{table}[t]
\caption{Numerical implementation of finite integrals with VEGAS~\cite{Lepage:1977sw,Hahn:2004fe}. The external energies are fixed to $p_{i,0} = 1$ in arbitrary units. The uncertainties are statistical.
\label{tab:numerics}}
\begin{tabular}{lr} \hline
& VEGAS \\ \hline
$\add{1}{d}{1,1,1} (r_1=r_1'=1)$ & $2.604161 (5) 10^{-3}$ \\
$\add{1}{d}{1,1,1} (r_1=r_1'=\qon{1})$ & $3.570478 (2) 10^{-3}$ \\ \hline
$\add{2}{d}{1,1,1,0,1,1,1}$ & $2.11138 (6) 10^{-6}$ \\ 
$\add{2}{d}{1,1,1,1,1,0,1}$ & $
9.2391 (2) 10^{-7}$ \\  \hline
$\add{3}{d}{1,\cdots,9}$ & $5.1555 (4) 10^{-9}$ \\
$\add{3}{d}{1,\cdots,9} (x_{89})$ & $2.25479 (4) 10^{-9}$ \\ \hline
$\add{4}{d}{1,\cdots,12}$ & $1.6207 (4) 10^{-11}$ \\
$\add{4}{d}{1,\cdots,12} (x_{10\cdots 12})$ & $1.63358 (6) 10^{-12}$ \\ \hline
$\add{5}{d}{1,\cdots,15}$ & $5.878 (5)  10^{-14}$ \\ 
$\add{5}{d}{1,\cdots,15} (x_{12\cdots 15})$ & $9.1271 (4) 10^{-16}$ \\
\hline
\end{tabular}
\end{table}

\section{Conclusions}
Infrared and threshold singularities are explicitly encoded in LTD through causal propagators that correspond to momentum configurations where particles propagate in definite directions. This manifest representation makes LTD particularly well suited to a streamlined, integrand-level analysis of the origin of such singularities. Leveraging this property, a new methodological strategy has been introduced in which finite integrands are obtained from a numerator Ansatz by requiring that the residues on the singular causal propagators vanish. This strategy reproduces results comparable to other approaches, but in a more compact and systematic manner, including loop pattern with raised propagators or defined in shifted spacetime dimensions. However, numerators typically induce a rapid UV scaling, which restricts the class of finite integrands that can be constructed in this way. To overcome this limitation, an intrinsically LTD-based construction with improved UV behaviour has been developed, offering better prospects for defining a basis of finite integrands at high loop orders.

\section{acknowledgments}
This work is supported by the Spanish Government and ERDF/EU - Agencia Estatal de Investigaci\'on MCIN/AEI/10.13039/501100011033,  Grants No. PID2023-146220NB-I00, No. EUR2025-164820, and No. CEX2023-001292-S. 
The work of KP is funded by Grant No. CEX2023-001292-S.
The work of JRS is funded by AEI, Grant No. PREP2023‐001474. 
The work of ST is supported by Generalitat Valenciana, Grant No. CIAPOS/2024/469. PKD is grateful to the European Commission MSCA Action COLLINEAR-FRACTURE, Grant Agreement No.~101108573, during which most of this project's work was completed. SRU acknowledges support from SECIHTI through the SNII programme.
This research was also supported by the Munich Institute for Astro-, Particle and BioPhysics (MIAPbP) which is funded by the Deutsche Forschungsgemeinschaft (DFG, German Research Foundation) under Germany's Excellence Strategy – EXC-2094 – 390783311.

\appendix
\section{Momentum labeling of ladder integral}
\label{app:ladder}

In this Appendix, we provide a detailed description of the loop labeling used for the numerical implementation of the multiloop ladder diagram shown in Fig.~\ref{fig:ladder_triangle}. The specific assignments of the loop three-momenta are as follows
\bea
&& \qb_1 =\lb_1+\pb_1~, \qquad
\qb_{s} = \lb_{s-1}+\pb_{12}~,\nn \\ &&
\qb_{\Lambda+s} = \lb_{s-1}~, \qquad ~
\qb_{2\Lambda+r} = \lb_{r-1, \bar r}~, \nn \\ &&
s \in [2, \Lambda+1]~, \quad r \in [2, \Lambda]~,
\eea
where $\Lambda$ is the number of loops, $\lb_{r-1,\bar r} = \lb_{r-1} - \lb_r$,
and in the centre of mass frame $\pb_{12} = \pb_1+\pb_2 = 0$. Therefore, $\qb_{s} = \qb_{\Lambda+s}$. We assume that $\pb_1$ is along the $z$-axis, and the loop three-momenta are parametrized as 
\beq
\lb_s = |\lb_s| (\sin \theta_s \cos \phi_s, \sin \theta_s \sin \phi_s, \cos \theta_s)~.
\eeq
Specifically, we use the change of variables
\beq
|\lb_s| = \frac{x_s}{1-x_s}~, \quad \cos \theta_s = 1-2v_s~, 
\eeq
with $x_s,v_s \in [0,1]$, and $\sin \theta_s = 2\sqrt{v_s(1-v_s)}$.

The LTD representation of the finite and threshold-free integrand representing the multiloop configuration in Fig.~\ref{fig:ladder_triangle} is
\bea
\add{\Lambda}{d}{1,\cdots, 3\Lambda} &=& \frac{1}{x_{1\cdots 3\Lambda}} \frac{1}{\lambda_{1,\Lambda+2;1} \lambda_{\Lambda+1;2\Lambda+1;12}} \nn \\ &\times&
\left( \prod_{s=3}^{\Lambda+1} \frac{1}{\lambda_{1, 2\Lambda+2, \cdots, 2\Lambda+s-1, \Lambda+s;1}}\right) \nn \\ &\times& \left(\prod_{r=2}^\Lambda \frac{1}{\lambda_{r, 2\Lambda+r, \cdots 3\Lambda, 2\Lambda+1;12}} \right)~.
\label{eq:ladder}
\eea

Soft singularities are absent in the integral of \Eq{eq:ladder}, even though the factor $x_{1\cdots 3\Lambda}$ can vanish for soft on-shell energies. This occurs because the soft singularities in the integrand are suppressed by the integration measure, which scales as $\lb_s^2$. However, this suppression becomes less apparent when the number of on-shell energies exceeds the number of loop momenta, requiring a multichanneling approach for numerical implementations. At two loops, VEGAS stills handles the soft singularities of the integrand effectively. In fact, implementing multichannelling yields only a minimal improvement. However, unstable numerical results are obtained at higher loop orders. 

The easiest way to scape multichanneling is to consider a numerator of the form $x_{2\Lambda+2 \cdots 3\Lambda}$, i.e. an integrand defined as 
\beq
\add{\Lambda}{d}{1,\cdots, 3\Lambda} (x_{2\Lambda+2 \cdots 3\Lambda}) = 
x_{2\Lambda+2 \cdots 3\Lambda} \add{\Lambda}{d}{1,\cdots, 3\Lambda}~.
\eeq
This integrand is still UV finite and quite stable numerically.

A numerical implementation of \Eq{eq:ladder} accounting for all possible soft channels would be quite complex and inefficient. A more streamlined approach requires a minimal partial fraction decomposition to manage soft singularities effectively. The idea is to separate the factor $1/x_{1\cdots 3\Lambda}$ into terms with less on-shell energies, such that the potential soft singularities of each term is effectively suppressed by the integration measure using different momentum parametrizations. Specifically, we identify the following transformations at three loops
\beq
\frac{1}{x_{1\cdots 9}} = \frac{1}{(x_{3467}+x_{1589}) x_2} \left( \frac{1}{x_{3467}} + \frac{1}{x_{1589}} \right)~,
\eeq
four loops 
\bea
\frac{1}{x_{1\cdots 12}} &=& \frac{1}{(x_{345789}+p_{1,0} x_{16} x_{10\cdots 12}) x_2} \nn \\ &\times& \left( \frac{p_{1,0}}{x_{345789}} + \frac{1}{x_{16}  x_{10\cdots 12}} \right)~,
\eea
and five loops
\bea
\frac{1}{x_{1\cdots 15}} &=& \frac{1}{(x_{3\cdots 6} x_{8\cdots 11}+p_{1,0}^2 x_{17} x_{12\cdots 15}) x_2} \nn \\ &\times& \left( \frac{p_{1,0}^2}{x_{3\cdots 6} x_{8\cdots 11}} + \frac{1}{x_{17} x_{12\cdots 15}} \right)~.
\eea

\bibliography{Finite_Integrals}

\end{document}